\RequirePackage[table]{xcolor}
\documentclass[11 pt]{article}
\usepackage[english]{babel}
\usepackage[utf8]{inputenc}
\usepackage{johd}
\usepackage{amsmath}
\usepackage{amsfonts}       % blackboard math symbols
\usepackage{amssymb}        % additional math symbols

\usepackage{subcaption}
\usepackage{multirow}
\usepackage{pifont}
\usepackage{algorithm}
\usepackage{algpseudocode}

\usepackage{xcolor}
\usepackage{hyperref}

\definecolor{maroon}{RGB}{177, 27, 87}
\definecolor{urlcolor}{rgb}{0.21,0.49,0.74}

\hypersetup{
    colorlinks=true,
    citecolor=maroon,
    linkcolor=black,
    urlcolor=maroon
}

\newcommand{\cmark}{\textcolor{green!60!black}{\ding{51}}}
\newcommand{\xmark}{\textcolor{red!70!black}{\ding{55}}}

\definecolor{gray_red}{RGB}{240,80,80}

\usepackage{wrapfig, sidecap, adjustbox, float}
\usepackage{booktabs}

\def\x{{\mathbf x}}

\def\y{{\mathbf y}}
\def\A{{\mathbf A}}
\def\E{{\mathbf E}}
\def\I{{\mathbf I}}

\usepackage{titling}

\pretitle{\centering\LARGE\bfseries}
\posttitle{\par\vskip 0.5em}

\title{Consistency Models for Fast MRI Reconstruction Using Regularization by Denoising}

\author{Merve~G\"ulle$^*$, Junno~Yun$^*$, Ya\c{s}ar Utku Al\c{c}alar, Mehmet~Ak\c{c}akaya \\\\
        \small Department of Electrical and Computer Engineering, University of Minnesota, MN, USA \\
        \small Center for Magnetic Resonance Research, University of Minnesota, MN, USA \\
        \small \texttt{\{glle0001, yun00049, alcal029, akcakaya\}@umn.edu}
}
\date{}

\begin{document}

\begingroup
\renewcommand{\thefootnote}{*}
\footnotetext{Merve G\"ulle and Junno Yun contributed equally to this work.}
\endgroup

\maketitle
\makeatletter
\typeout{CURRENT FONT SIZE = \f@size pt}
\makeatother
\begin{abstract} 
Diffusion models (DMs) have emerged as powerful generative priors for MRI reconstruction with promising results. Yet DM-based methods require extensive iterative refinement, limiting their practical deployment. Consistency models (CMs) provide a compelling alternative, aiming to map out the diffusion trajectory in a single pass, enabling faster generation. In this work, we propose \emph{CM-RED}, a novel MRI reconstruction method that integrates a pretrained CM into the regularization by denoising (RED) scheme. Our method builds on accelerated proximal gradient RED (RED-APG), and further incorporates controlled noise injection during the update steps to enhance generative diversity and accelerate convergence. Extensive experiments on the fastMRI knee and brain datasets demonstrate that CM-RED achieves high-quality reconstructions across multiple anatomies, contrast weights, acceleration factors, and undersampling patterns, using only 4 network function evaluations (NFEs). The proposed method consistently outperforms existing DM- and CM-based approaches in both quantitative metrics and visual fidelity, and exhibits strong robustness to hyperparameter variations, highlighting CM-RED as an efficient and effective generative framework for accelerated MRI reconstruction. The source code and pretrained models are publicly available at \url{https://github.com/MerveGulle/CM-RED}.
\end{abstract}

\noindent

\section{Introduction}\label{sec:introduction}

Image reconstruction from accelerated MRI acquisitions has seen steady progress over the past decades ~\cite{pruessmann1999sense, lustig2007sparse, hammernik2018VarNet, akcakaya2022reconbook}. Nonetheless, due to the ill-posed nature of this inverse problem at high acceleration rates, conventional reconstruction techniques such as parallel imaging~\cite{pruessmann1999sense, griswold2002grappa} and compressed sensing (CS)~\cite{lustig2007sparse} struggle to simultaneously preserve image fidelity and suppress artifacts~\cite{hammernik2023SPM}. More recently, machine/deep learning (ML/DL) methods have been adopted, including data-driven approaches that learn direct mappings~\cite{wang2016accelerating,lee2018deep,han2019k, akcakaya2019RAKI}, and physics-driven deep learning (PD-DL) methods \cite{hammernik2018VarNet, schlemper2018deep, aggarwal2019MoDL, knoll2020deep-survey, saberi2026umpire}. The latter, encompassing unrolled networks~\cite{aggarwal2019MoDL, hammernik2018VarNet, schlemper2018deep} and  plug-and-play (PnP) reconstruction~\cite{ahmad2020plug}, 
incorporate the measurement operator into the reconstruction via data consistency operations along with learned regularization modules. In particular, unrolled network have shown improved performance over data-driven approaches~\cite{muckley2021fastMRIchallenge-2}.

Despite their success, these approaches have limitations. Unrolled networks, are typically trained for specific imaging configurations, including anatomies, sampling patterns, acceleration factors, and image contrasts, which limits their performance in out-of-distribution scenarios~\cite{yaman2022zeroshot}. On the other hand, while PnP methods offer improved flexibility, they suffer from a distribution mismatch since the intermediate reconstructions contain structured artifacts instead of simple Gaussian noise, leading to performance degradation~\cite{shastri2022expectation}. Furthermore, the denoisers are optimized for low-noise degradation, limiting applications in severely ill-posed regimes~\cite{park2026stochastic}.

Recently, generative models have emerged as a promising paradigm for inverse problems, including MRI reconstruction. In particular, diffusion models (DM)~\cite{ho2020ddpm, song2021scoreSDE, karras2022elucidating} have demonstrated a strong ability to model complex image distributions. In the context of MRI reconstruction, diffusion-based sampling frameworks~\cite{jalal2021robust, song2022solving-medical, chung2024decomposed, alcalar2024ZAPS} exhibit robustness to variations in acquisition protocols and data distributions. However, their reliance on hundreds of sequential sampling steps leads to substantial computational overhead, limiting their applicability in time-sensitive workflows and large-scale deployment scenarios~\cite{webber2024diffusion}. Motivated by the need for efficient generative priors, consistency models (CM)~\cite{song2023_consistency_models} have recently been proposed in the broader image generation field as efficient models for sampling the DM trajectory in a single step while retaining the expressive power of the learned image prior.

In this paper, we propose \textbf{CM-RED}, a novel framework that leverages a pretrained CM as a proximal prior within the RED formulation~\cite{romano2017little}. Our main contributions are: 
\begin{itemize}
    \item We propose CM-RED, a plug-and-play framework that integrates a CM as a learned proximal prior within the RED formulation, enabling high-quality reconstruction of accelerated MRI measurements with only a few iterations.
    \item To further utilize the fast generative sampling properties of CMs, we refine the accelerated proximal gradient scheme (RED-APG)~\cite{reehorst2018regularization} framework by incorporating controlled noise injection and iteration-adaptive denoising strength for CMs. This design accelerates convergence, enhances robustness of reconstructions across a wide range of noise conditions, and synergistically utilizes the powerful generative capabilities of CMs.
    \item We establish unified hyperparameters across anatomies, contrasts, and sampling patterns, with settings differing only by acceleration rate, and demonstrate robustness to moderate parameter perturbations.
    \item We validate CM-RED through extensive evaluations on fastMRI knee and brain datasets. We also train CMs from scratch for various anatomies. Our method consistently outperforms existing DM- and CM-based inverse problem solvers, while providing a $25\times$--$250\times$ speed-up.
\end{itemize}

Preliminary versions of this work were presented in a conference publication~\cite{gulle2026_ISBI} and abstract~\cite{gulle_CM_RED_ISMRM}. The current work substantially extends these through broader evaluations across knee and brain MRI, multiple contrasts, acceleration factors, and undersampling patterns, together with an expanded methodological description, additional ablation, and comprehensive comparisons with DM- and CM-based reconstructions.

\section{Background and Related Works}
\label{sec:background}

\subsection{MRI Reconstruction Inverse Problem}
MR image reconstruction solves:
\begin{equation}
\label{eq:map_opt}
\arg\min_{\x} \left\| \y - \E_{\Omega}\x \right\|_2^2 + \mathcal{R}(\x),
\end{equation}
where $\y \in \mathbb{C}^M$ is the acquired k-space measurements,  $\mathbf{E}_{\Omega} \in \mathbb{C}^{M \times N}$ is the multi-coil encoding operator that samples locations indexed by $\Omega$, $\mathbf{n} \in \mathbb{C}^M$ is i.i.d. Gaussian noise. Here, the first term enforces data fidelity with the acquired data, and
$\mathcal{R}(\cdot)$ is a regularizer.

In classical parallel imaging, this problem is solved either without a regularizer or with a Tikhonov-type regularizer, leading to a linear reconstruction process~\cite{pruessmann1999sense, pruessmann2001cgsense}. Subsequently, sparsity-focused $\ell_1$-norm based regularizers were popularized with compressed sensing~\cite{lustig2007sparse}. These lead to a non-linear optimization problem, which is typically solved with iterative algorithms that alternate between data fidelity updates and proximal operations associated with $\mathcal{R}(\x)$~\cite{fessler2020SPM}. 

More recently, ML/DL approaches have been popular to design powerful regularizers or proximal operators, parameterized by neural networks and learned directly from data. Among these methods, algorithm unrolling~\cite{hammernik2018VarNet, aggarwal2019MoDL, hammernik2023SPM} follows the alternating optimization principle, but learns both the data fidelity and regularization modules for a fixed number of unrolls. While such end-to-end learned methods achieve high reconstruction quality on in-distribution data, they often exhibit reduced robustness to variations in sampling patterns, acceleration factors, or SNR, limiting their generalizability~\cite{yaman2022zeroshot}. PnP methods~\cite{ahmad2020plug} generalize this framework by replacing the explicit regularizer $\mathcal{R}(\x)$ with a pre-trained denoiser that acts as an implicit proximal operator. However, PnP performance may be limited since denoisers are typically optimized for Gaussian noise, whereas the errors encountered during iterative updates exhibit complex, non-Gaussian behavior~\cite{shastri2022expectation}, which may degrade reconstruction quality.

\subsection{Diffusion Models}
DMs have achieved remarkable success in data synthesis across a wide range of applications~\cite{ho2020ddpm, song2021scoreSDE, karras2022elucidating}. 
Among them, Denoising Diffusion Probabilistic Models (DDPMs)~\cite{ho2020ddpm} generate data by progressively denoising samples through a series of steps. In DDPMs, the forward diffusion process gradually corrupts the data by adding Gaussian noise over $T$ time steps:
\begin{equation}
    \x_t = \sqrt{1-\beta_t}\,\x_{t-1} + \sqrt{\beta_t}\,\boldsymbol{\epsilon}, 
    \quad \boldsymbol{\epsilon} \sim \mathcal{N}(0, \mathbf{I}),
\end{equation}
which yields the closed-form distribution:
\begin{equation}
    q(\x_t \mid \x_0) = \mathcal{N}\!\left(\x_t;\sqrt{\bar{\alpha}_t}\,\x_0,(1-\bar{\alpha}_t)\mathbf{I}\right),
\end{equation}
where $\alpha_t = 1-\beta_t$ and $\bar{\alpha}_t=\prod_{s=1}^t \alpha_s$, with $\{\beta_t\}_{t=1}^T$ denoting a predefined noise schedule. 
The reverse process aims to recover samples from the target data distribution by iteratively removing noise. 
A neural network $\boldsymbol{\epsilon}_\theta(\x_t,t)$ is trained to estimate the noise added during the forward diffusion, enabling the approximation of the reverse-time dynamics.
After training, sampling starts from a Gaussian noise sample $\x_T \sim \mathcal{N}(0,\mathbf{I})$ and proceeds backward from $t=T$ to $1$.

\vspace{1ex}
\noindent \textbf{{DM-based Inverse Problem Solvers.}}
From a Bayesian perspective, the objective in~\eqref{eq:map_opt} for i.i.d. Gaussian noise can be interpreted as maximum a posteriori (MAP) estimation
\begin{equation}
    \arg \max_{\bf x} \,  \log p(\y \mid \x) \, + \log p(\x). 
\label{eq:bayes_posterior}
\end{equation}
Utilizing the connection between the score function and DMs, several inverse problem solvers have been proposed to  combine these generative priors with physics-based data fidelity terms~\cite{jalal2021robust,chung2023dps, chung2022scoreMRI, song2022solving-medical, chung2024decomposed}. In particular, these rely on 
\begin{equation}
    \nabla_{\x_t} \log p(\x_t|\y) = \nabla_{\x_t} \log p(\x_t) + \nabla_{\x_t} \log p(\y|\x_t),
    \label{eq:score_posterior}
\end{equation}
where the first unconditional score term is provided by a pre-trained DM, while the second term, the likelihood score, which is crucial for enforcing data fidelity to the physical measurements $\y$, is intractable~\cite{chung2023dps, song2023pgdm}. Early DM-based MRI solvers~\cite{jalal2021robust, luo2023bayesian, chung2022scoreMRI} circumvented this issue with heuristic approximations that directly evaluate the physical forward model on the noisy intermediate state:
\begin{equation}
    \nabla_{\x_t} \log p(\y|\x_t) \approx \nabla_{\x_t} \log p(\y|\x_0)\big|_{\x_0=\x_t} \propto  - \A^H(\A\x_t - \y).
\end{equation}
This approximation was combined with either gradient-based Langevin-type samplers~\cite{jalal2021robust, luo2023bayesian} or numerical SDE solvers and a strict data consistency step~\cite{chung2022scoreMRI}. 

On the other hand, more recent methods, such as Diffusion Posterior Sampling (DPS)~\cite{chung2023dps} and Denoising Diffusion Samplers (DDS)~\cite{chung2024decomposed} note that measurements ${\bf y}$ is connected via ${\bf A}$ only to the clean data manifold, and provide new approximations for the likelihood term instead of relying on heuristic projections on noisy latents. Specifically, DPS evaluates this approximate measurement likelihood term at the Tweedie estimate of the clean signal given as:
\begin{equation}
    \hat{\x}_0
    =
    \left(
    \x_t - \sqrt{1-\bar{\alpha}_t}\,\boldsymbol{\epsilon}_\theta(\x_t,t)
    \right) /\sqrt{\bar{\alpha}_t},
    \label{eq:tweedie_est}
\end{equation}
showing improved performance across a range of inverse problems~\cite{chung2023dps}. However, it typically requires hundreds of NFEs to achieve stable convergence, limiting its applicability in large-scale imaging applications. Building on DPS, DDS makes the observation that the clean data distribution can be locally approximated by an affine subspace corresponding to the tangent space of the data manifold. It uses this local linear structure to replace the single gradient correction in DPS with a few conjugate gradient (CG) iterations restricted to the subspace that spans the local tangent of the manifold, thereby reducing the NFEs to $\sim$100. Nevertheless, despite this improvement, the computational burden of DDS remains substantial, rendering deployment still impractical. 

\begin{figure}[t]
  \centering
  \includegraphics[width=0.9\columnwidth]{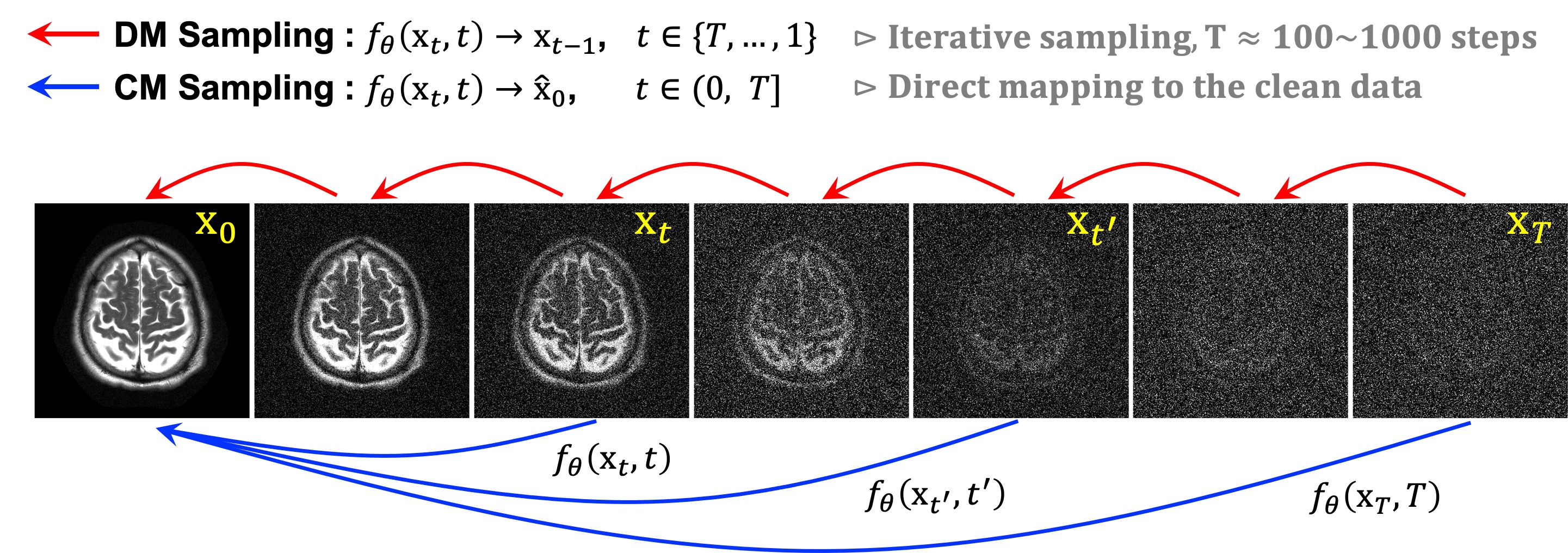}
  \caption{Illustration of DM and CM trajectories on MRI data.}
  \vspace{-3pt}
  \label{fig:DM_vs_CM}
\end{figure}

\subsection{Consistency Models}
CMs~\cite{song2023_consistency_models} offer an attractive alternative to DMs by achieving high-quality generation with a single or few NFEs, thereby alleviating the heavy sampling burden of DMs. The key distinction between DMs and CMs lies in their denoising mechanisms: DMs perform sequential denoising along the entire diffusion trajectory (red arrows in Fig.~\ref{fig:DM_vs_CM}), whereas CMs are trained to directly map a noisy sample at any timestep $t$ to a clean estimate in a single step~\cite{song2023_consistency_models} (blue arrows in Fig.~\ref{fig:DM_vs_CM}). This is achieved by defining a consistency function
\begin{equation}
    f_{\theta}(\mathbf{x}_t, t) = f_{\theta}(\mathbf{x}_{t'}, t'), \quad \forall\, t, t' \in [\epsilon, T].
\end{equation}
with a boundary condition $f_{\theta}(\mathbf{x}_{\epsilon}, \epsilon) = \mathbf{x}_{\epsilon}$. The model parameters $\theta$ are optimized using a consistency objective that enforces agreement between outputs corresponding to two noisy versions of the same underlying clean sample:
\begin{equation}
    \mathcal{L}_{\mathrm{CM}} =
    \mathbb{E}_{\mathbf{x}_0, t, t'} \!\left[
        w(t)\, d\!\left(f_{\theta}(\mathbf{x}_t, t),\, f_{\theta^-}(\mathbf{x}_{t'}, t')\right)
    \right],
\end{equation}
where $w(t)$ is a weighting function, $d(\cdot,\cdot)$ is a distance metric such as $\ell_2$, and $f_{\theta^-}$ is a teacher network. When a CM is obtained via distillation from a pretrained DM, the sample $\mathbf{x}_{t'}$ is generated from $\mathbf{x}_t$ using an ordinary differential equation (ODE) solver~\cite{song2023_consistency_models}. As a result, the CM learns to recover the clean sample from any point along the ODE trajectory. This training paradigm fundamentally differs from the denoising objective of DMs, which focuses on learning an average denoising direction rather than a globally consistent mapping.\vspace{5pt}

\vspace{1ex}
\noindent \textbf{{CM-based Inverse Problem Solvers.}}
Owing to their computational efficiency and expressive learned priors, CMs have recently attracted attention for inverse problem solving. Most existing CM-based approaches~\cite{zhao2024cosign, garber2025CM4IR}, however, were developed in the context of natural image restoration, and their applicability to MRI reconstruction remains limited. The original CM framework~\cite{song2023_consistency_models} incorporates data fidelity during the sampling process, yet still requires on the order of 40 NFEs, rendering it impractical for large-scale imaging problems. CoSIGN~\cite{zhao2024cosign} augments this framework with a ControlNet that explicitly encodes the forward operator, enabling guidance in a few inference steps but requires retraining for each acquisition setting. CM4IR~\cite{garber2025CM4IR} enforces measurement consistency via a pseudo-inverse back-projection step, effectively combining CMs with linear reconstruction techniques. Nevertheless, this approach does not readily generalize to MRI forward models, where pseudo-inverse solutions $\E_\Omega^{\dagger}\y$ typically suffer from residual aliasing artifacts and noise amplification~\cite{pruessmann2001cgsense}.

\section{Methods}
\label{sec:methods}

\subsection{Regularization by Denoising (RED)}
Regularization by Denoising (RED) is a framework to incorporate image denoisers into inverse problem solvers through an explicit regularization functional. Similar to plug-and-play (PnP) methods, RED leverages powerful pretrained denoisers within iterative reconstruction algorithms, while additionally providing an explicit regularization term. Specifically, given a denoising operator $D(\cdot)$, RED defines the regularization term:
\begin{equation}
\mathcal{R}_{\mathrm{RED}}(\x) = \frac{\lambda}{2}\,\x^\top \big(\x - D(\x)\big),
\end{equation}
where $\lambda > 0$ controls the strength of regularization. Under certain assumptions on the denoiser, such as local homogeneity and Jacobian symmetry, the gradient of $\mathcal{R}_{\mathrm{RED}}(\x)$ admits a closed-form expression proportional to $\x - D(\x)$, which allows RED to be naturally embedded into gradient-based optimization schemes~\cite{romano2017little}.

Several variants of RED has been proposed for integrating denoisers into iterative reconstruction algorithms. Among these, a quadratic majorization of the RED objective has been explored in~\cite{reehorst2018regularization}, leading to the RED proximal gradient (RED-PG) algorithm and its accelerated variant, RED-APG. These methods decouple the data fidelity and denoising updates, and the accelerated version further incorporates a momentum term following the data fidelity update, which has been shown to substantially reduce the number of iterations required for convergence in practice~\cite{reehorst2018regularization}.

\subsection{Proposed Method: CM-RED}
\label{sec:cm-red}

To leverage the fast generative sampling capabilities of consistency models (CMs) for MRI reconstruction, we propose CM-RED, a reconstruction framework that integrates a pretrained CM within the RED-APG formulation. Specifically, we employ the CM sampling as the surrogate denoising operator in this optimization framework, combining CM-based prior updates with explicit data fidelity enforcement and momentum acceleration. Notably for this to work efficiently, we incorporate controlled noise injection into the CM updates to enable stronger corrective denoising steps during reconstruction. The overall process for CM-RED is summarized in Alg.~\ref{alg:cm-red}, where $N$ denotes the number of iterations, indexed in reverse order from $N-1$ to $0$. Next, we detail the individual components. \vspace{4pt}

\vspace{1ex}
\noindent \textbf{{CM-based Proximal Update with Noise Injection.}} At each iteration, CM-RED first applies a CM-based prior update to the previous iterate $\x_{n+1}^{(m)}$. Specifically, the pretrained CM $f_\theta(\cdot)$ produces a refined reconstruction from a perturbed version of the current iterate, and the resulting CM output is combined with the previous estimate through a weighted interpolation (Alg.~\ref{alg:cm-red}, line 4):
\begin{equation}
\tilde{\x}_n
=
\nu_n f_\theta \big(\x_{n+1}^{(m)} + \sigma_n \epsilon,\tilde{\sigma}_n \big)
+
(1-\nu_n)\x_{n+1}^{(m)},
\end{equation}
where $\epsilon \sim \mathcal{N}(0,\mathbf{I})$, $\sigma_n$ is the injected noise standard deviation, $\tilde{\sigma}_n$ is the CM denoising noise level, and $\nu_n \in (0,1)$ controls the contribution of the CM output. This interpolation allows the influence of the CM output to gradually increase throughout the reconstruction process, enabling stronger CM-driven refinement in later iterations.

Notably, Gaussian perturbations are intentionally injected into the CM input before denoising at every iteration. CMs are trained to satisfy the boundary condition $f_{\theta}(\mathbf{x}_{\epsilon}, \epsilon) = \mathbf{x}_{\epsilon}$, implying that the CM increasingly behaves like an identity mapping as the denoising noise level approaches zero. Consequently, operating the CM at very low noise levels leads to only minor corrective updates within the iterative reconstruction process. On the other hand, injecting additional perturbations increases the amount of corruption that must subsequently be removed by the CM, thereby producing larger corrective denoising updates during reconstruction. In practice, these larger updates enable the reconstruction to reach high-quality solutions in substantially fewer iterations. Similar stochastic perturbation strategies have previously been explored in CM-based inverse problem solvers~\cite{song2023_consistency_models, garber2025CM4IR} and stochastic optimization frameworks~\cite{atchade2017perturbed, renaud2024plug, park2026stochastic}. The effect of noise injection on reconstruction quality and convergence behavior is further investigated empirically in Sec.~\ref{sec:ablation_studies}.

\begin{center}
\begin{minipage}{0.9\textwidth}
\begin{algorithm}[H]
\caption{Proposed Algorithm: CM-RED}
\label{alg:cm-red}
\begin{algorithmic}[1]
    \Statex \hspace*{-2em} \textbf{Require:} 
    CM $f_\theta(\cdot)$, 
    encoding operator $\E_{\Omega}$, 
    number of iterations $N$, 
    momentum coefficients $\mu_n$, 
    regularization parameter $\lambda_n$, 
    CM output scales $\nu_n$, 
    injected noise levels $\sigma_n$, 
    CM denoising levels $\tilde{\sigma}_n$.

    \vspace{+0.15cm}
    \Statex {\textit{\color{gray_red} $\triangleright$ Initialize with CG-SENSE reconstruction}}
    
    \State $\x_N = \x_{N}^{\scriptscriptstyle (m)} = (\E_{\Omega}^{H}\E_{\Omega})^{-1} \E_{\Omega}^{H} \y$
    \vspace{+0.15cm}
    
    \For{$n = N-1$ \textbf{to} $0$}
        \vspace{+0.1cm}
        \Statex \hspace*{+1.1em} {\textit{\color{gray_red} $\triangleright$ CM-based proximal update with noise injection}}
        \State $\epsilon \sim \mathcal{N}(\mathbf{0}, \mathbf{I})$
        \State $\tilde\x_n =
            \nu_n \, f_\theta\Big(
                \x_{n+1}^{\scriptscriptstyle (m)} + \sigma_n \, \epsilon, \tilde{\sigma}_n
            \Big)
            + (1-\nu_n) \, \x_{n+1}^{\scriptscriptstyle (m)}$

        \vspace{+0.2cm}
        \Statex \hspace*{+1.1em} {\textit{\color{gray_red} $\triangleright$ Data-fidelity (DF) update via CG}}
        \State $\x_n =
            (\E_{\Omega}^{H}\E_{\Omega} + \lambda_n \I)^{-1} \Big(
                \E_{\Omega}^{H} \y + \lambda_n \tilde\x_n
            \Big)$

        \vspace{+0.2cm}
        \Statex \hspace*{+1.1em} {\textit{\color{gray_red} $\triangleright$ Momentum acceleration}}
        \State $\x_n^{\scriptscriptstyle (m)} = \x_n + \mu_n \, (\x_n - \x_{n+1})$
        
        \vspace{+0.15cm}
    \EndFor
    
    \State \textbf{Return:} $\x_0$
\end{algorithmic}
\end{algorithm}
\vspace{-6pt}
\end{minipage}
\end{center}

\vspace{1ex}
\noindent \textbf{{Data Fidelity.}}  Following the CM-based proximal update, data fidelity is enforced by solving the following quadratic penalized least-squares problem:
\begin{equation}
\x_n = \arg \min_{\x}\|\E_{\Omega}\x-\y\|_2^2 + \lambda_n\|\x-\tilde{\x}_n\|_2^2,
\end{equation}
where $\tilde{\x}_n$ corresponds to the CM-refined reconstruction estimate. This update enforces consistency with the measured data while allowing the CM to provide strong generative regularization. The resulting linear system as given in Alg.~\ref{alg:cm-red} line 5, is solved efficiently using CG. 

\begin{figure}[b]
  \centering
  \includegraphics[width=0.9\columnwidth]{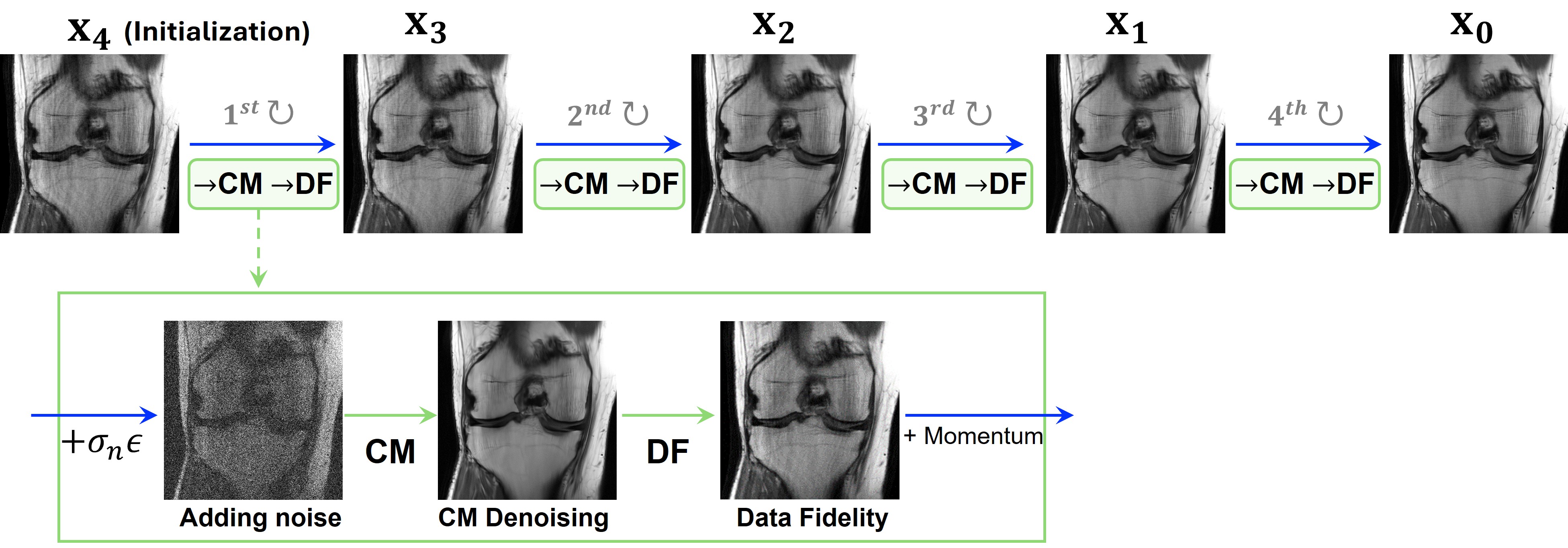}
  \caption{Illustration of CM-RED steps with $N\!=\!4$, as described in Alg.~\ref{alg:cm-red}.}
    \vspace{-3pt}
  \label{fig:cm_red_process}
\end{figure}

\vspace{1ex}
\noindent \textbf{{Momentum Acceleration.}} Momentum acceleration (Alg.~\ref{alg:cm-red}, line 6), inherited from RED-APG, is applied after the data fidelity update according to
\begin{equation}
\x_n^{(m)} = \x_n + \mu_n(\x_n - \x_{n+1}),
\end{equation}
where $\mu_n$ denotes the momentum coefficient. The effect of momentum on reconstruction quality and convergence behavior is further investigated empirically in Sec.~\ref{sec:ablation_studies}.

A visualization of the CM-RED reconstruction process for 4 iterations is provided in Fig.~\ref{fig:cm_red_process}, depicting the progressive refinement of the intermediate images. 

\begin{table}[t]
\caption{Complete set of hyperparameters for our four-step sampling setup, including noise schedule, regularization parameters, CM out scale parameter, and momentum coefficients.}
\vspace{-4pt}
\label{tab:hyperparams}
\begin{center}
\begin{small}
\renewcommand{\arraystretch}{1.1}
\setlength{\tabcolsep}{10.0pt}

\begin{tabular}{ccccccc}

\toprule

\rowcolor{gray!20}
Acc. & $i_N$ & $\gamma$ &  $\delta_n$ & $\kappa$ & $\rho$ & $\mu^{(0)}$ \\
\midrule

\addlinespace[2pt]
4$\times$ & 50 & 0.1 & $\big(0.4,\,3.0,\,3.0,\,2.5\big)$ &
$\big(0.5,\,5.0\big)$ & 
--2.0 & 
0.9  \\

\addlinespace[2pt]
8$\times$ & 50 & 0.1 & 
$\big(0.5,\,7.0,\,6.0,\,3.5\big)$ & 
$\big(1.0,\,2.5\big)$ & 
--3.0 & 
0.9  \\
\arrayrulecolor{gray!40}

\arrayrulecolor{black}\midrule

\end{tabular}
\end{small}
\end{center}
\vspace{-3ex}
\end{table}

\section{Experimental Setup}
\label{sec:exp_setup}

\subsection{Datasets}

We conducted a comprehensive evaluation of the proposed method, assessing its performance both quantitatively and qualitatively across multiple acceleration factors, undersampling patterns, and datasets. All experiments were performed on complex-valued multi-coil k-space measurements from the New York University (NYU) fastMRI dataset~\cite{knoll2020fastmri_dataset-journal}, acquired with institutional review board approval. The dataset includes fully-sampled coronal PD and PD-FS knee scans, as well as axial T1-, T2-, and FLAIR-weighted brain MRI scans.

\begin{figure}[!ht]
  \centering
  \includegraphics[width=1.0\textwidth]{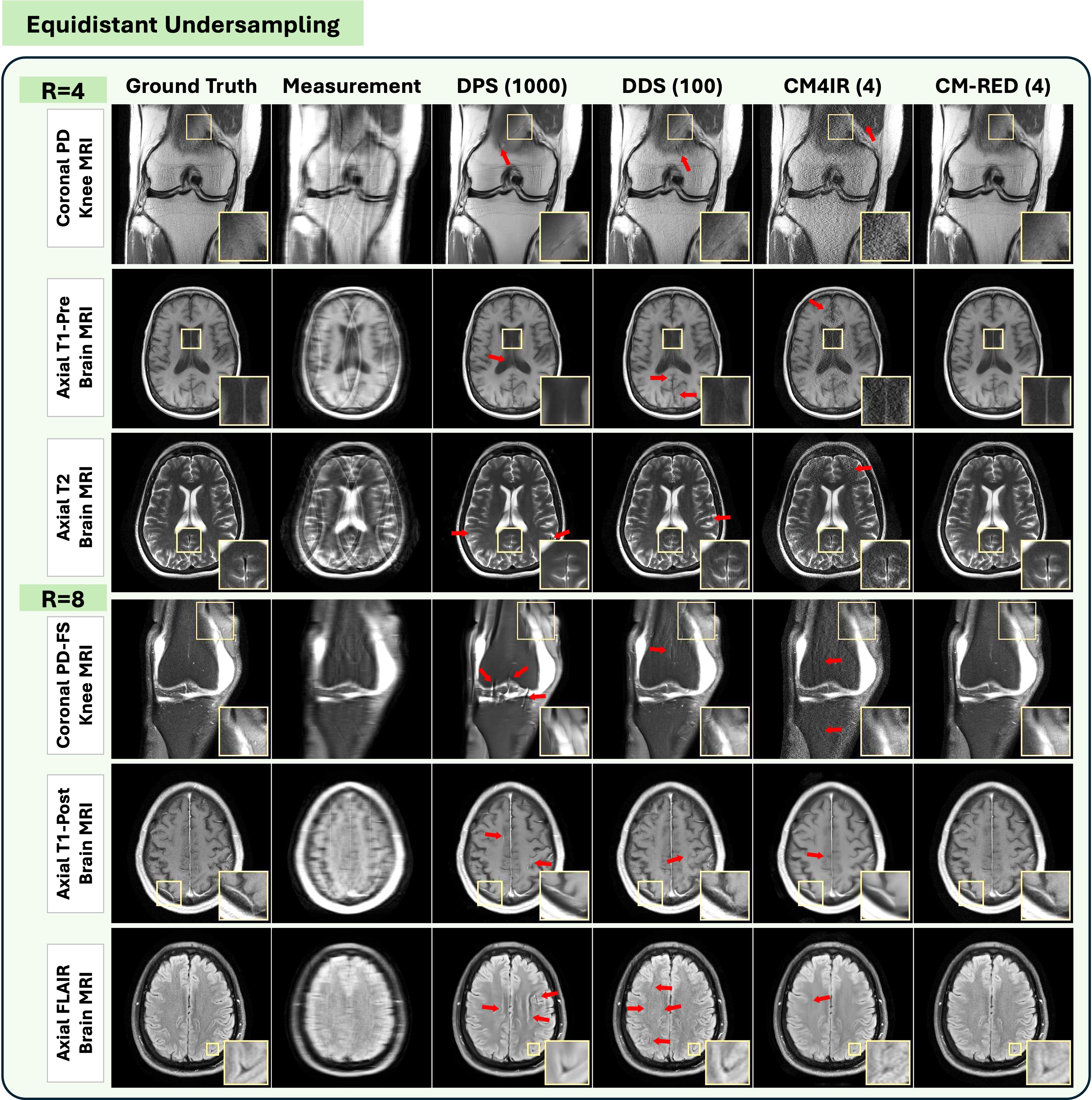}
    \caption{Representative reconstructions of DM- and CM-based methods on fastMRI knee and brain data with equidistant $R=4$ and $R=8$ undersampling. The number of NFEs used by each method is shown in parentheses. Yellow boxes indicate magnified regions, and red arrows highlight residual artifacts. Best viewed when zoomed in.}
  \label{fig:results_equi}
\end{figure}

\subsection{DM and CM Training Details}
For training on the knee dataset, we utilized 973 subjects from the fastMRI knee training set, excluding the first and last five slices of each volume, following the protocol in prior work~\cite{chung2024decomposed}. For the brain dataset, 4,469 subjects from the fastMRI brain training set were initially considered, and volumes with image width smaller than 320 were discarded, resulting in 4,267 subjects used for training. In all cases, the complex-valued coil-combined images were represented as two-channel inputs by concatenating the real and imaginary components along the channel dimension, resulting in a spatial resolution of $320 \times 320 \times 2$.

We used a pretrained CM for the knee dataset from our earlier work~\cite{gulle2026_ISBI}. For the brain dataset, the CM was trained from scratch. Following the publicly available implementations provided in~\cite{song2023_consistency_models, karras2022elucidating}, we trained EDM/CM for brain datasets in two stages with the same configurations in~\cite{zhao2024cosign, gulle2026_ISBI}. Specifically, the EDM model for the brain dataset was trained for 700K iterations with a batch size of 16, and then distilled into a CM over 1.05M iterations using a batch size of 12. All training was performed on NVIDIA A100 GPUs.

\begin{figure}[!ht]
  \centering
  \includegraphics[width=1.0\textwidth]{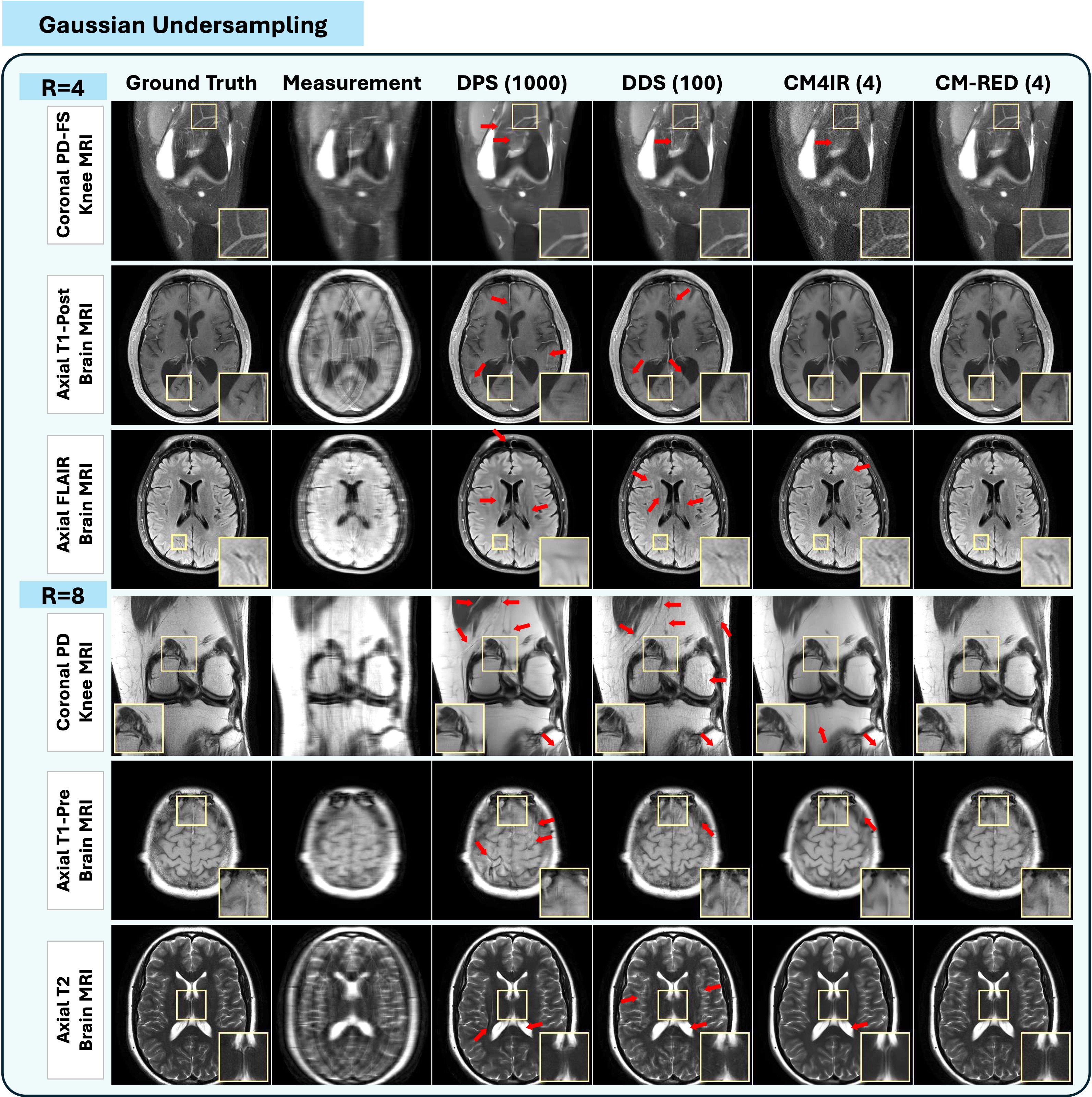}
  \caption{Representative reconstructions for Gaussian undersampling. Number of NFEs for by each method is provided in parentheses. Yellow boxes indicate magnified regions, and red arrows highlight residual artifacts. Best viewed zoomed in.}
\label{fig:results_gauss}
\end{figure}

\subsection{CM-RED Implementation Details} \label{sec:implementation_details}

CM-RED was implemented using $N=4$ outer iterations (except the ablation studies), with each data fidelity update solved using 10 CG iterations. The algorithm includes a small number of reconstruction hyperparameters that control the injected noise level, the CM denoising level, the relative contribution of the CM output, the quadratic penalty parameter, and the momentum coefficient. These parameters correspond to the quantities in Alg.~\ref{alg:cm-red}: $\sigma_n$, $\tilde{\sigma}_n$, $\nu_n$, $\lambda_n$, and $\mu_n$, respectively.

\vspace{1ex}
\noindent \textbf{{Noise Schedule.}} 
We employ a noise scheduling strategy that gradually decreases the injected noise across iterations using a small set of hyperparameters, similar to CM4IR~\cite{garber2025CM4IR}. Let $\{\beta_i\}_{i=1}^{T} \subset (0,1)$ denote the forward noising variance schedule in standard DDPM notation, with $\alpha_i = 1 - \beta_i$ and $\bar{\alpha}_i = \prod_{j=1}^{i} \alpha_j$. We first select an initial diffusion index $i_N \in [1,T]$, which determines the injected noise level at the first CM-RED iteration. To avoid overloading the diffusion index with the CM-RED iteration index, we define an auxiliary sequence $\{a_n\}_{n=0}^{N-1}$ as
\begin{equation}
    a_{N-1} = \bar{\alpha}_{i_N},
    \qquad
    a_n = \min\{(1+\gamma)a_{n+1},\,1\},
\end{equation}
where $\gamma>0$. The injected perturbation level at iteration $n$ is then given by
\begin{equation}
    \sigma_n = \sqrt{1-a_n},
    \qquad n=N-1,\ldots,0.
\end{equation}
Note $i_N$ controls the starting perturbation level, while $\gamma > 0$ indirectly controls how quickly the perturbation level decays over the CM-RED iterations.

\vspace{1ex}
\noindent \textbf{{CM Denoising Level.}} 
The CM denoising noise level $\tilde{\sigma}_n$ is allowed to differ from the injected perturbation level $\sigma_n$. Following the observation in prior CM-based inverse problem solvers~\cite{garber2025CM4IR, gulle2025consistency} that the denoising level can be chosen slightly larger than the injected perturbation level, we parametrize
\begin{equation}
    \tilde{\sigma}_n = (1+\delta_n)\sigma_n,
    \qquad n=N-1,\ldots,0,
\end{equation}
where $\delta_n \geq 0$ is an iteration-dependent offset. This parameter controls the effective denoising strength of the CM update at each iteration.

\vspace{1ex}
\noindent \textbf{{CM Output Scaling Parameter.}} 
At each iteration, the CM output is combined with the current momentum iterate through a $\nu_n$-weighted linear combination, following the RED-APG strategy of blending the denoiser-based update with the current reconstruction estimate. To simplify the tuning, we parameterize $\{\nu_n\}_{n=1}^N$ using a smooth, monotonically increasing schedule constrained to the interval $(0,1)$. The scaling coefficients are defined as 
$$s_n = \kappa_1 + \frac{n-1}{N-1}\,(\kappa_2 - \kappa_1), 
\quad \kappa_2 \ge \kappa_1, 
\quad \nu_n = 1 - e^{-s_n}$$
where $\kappa_1$ and $\kappa_2$ control the initial and final relative weighting in the CM output. This parametrization ensures that $\nu_n$ increases gradually approaches 1, enabling a controlled transition from data fidelity-dominated updates to CM-driven refinement while keeping hyperparameter tuning tractable. 

\vspace{1ex}
\noindent \textbf{{Data Fidelity Penalty Parameter.}} 
The parameter $\lambda_n$ controls the strength of the quadratic penalty that couples the reconstruction estimate to the CM-refined output $\tilde{\x}_n$ in the data fidelity update. In our implementation, we use a constant value across iterations, i.e., $\lambda_n = \lambda$, to avoid introducing additional iteration-dependent hyperparameters. This choice follows the common practice of sharing reconstruction parameters across iterations in unrolled MRI reconstruction frameworks \cite{aggarwal2019MoDL} to reduce parameter complexity. To ensure positivity and facilitate tuning over a wide range of small positive penalty values, $\lambda$ is parameterized using a softplus transformation, 
$$\lambda = \text{softplus}(\rho) = \log(1 + e^{\rho}),$$
where $\rho\in\mathbb{R}$ is the scalar parameter being tuned.

\vspace{1ex}
\noindent \textbf{{Momentum Coefficient.}} 
Momentum acceleration is applied after the data fidelity update, following the RED-APG update structure. In our implementation, we use a simple decreasing momentum schedule. Starting from an initial value $\mu^{(0)}$, the momentum parameter is reduced at each iteration as $$\mu_n = \mu^{(0)} / (N-n).$$ As a result, momentum updates are gradually dampened, limiting oscillatory behavior in the final stages of the algorithm.

The complete set of hyperparameters employed in each experiment is summarized in Tab.~\ref{tab:hyperparams}. As shown, the well-tuned hyperparameters are shared across all experiments, with the exception of those distinguishing acceleration factors $R=4$ and $R=8$. The same configuration was used across anatomies, contrast weightings, and undersampling patterns, with separate settings only for acceleration rates.

\subsection{Evaluations and Comparison Methods}
For evaluation, the corresponding fully sampled k-space data were retrospectively undersampled using uniform (equidistant) and Gaussian random masks at acceleration factors of $R \in \{4, 8\}$. For all cases, 24 central k-space lines were retained, except for Gaussian random sampling at $R = 8$, where 12 central lines were preserved. We selected 10 subjects from the validation set, resulting in 228 PD, 240 PD-FS, 157 axial T1 (pre-contrast), 158 axial T1 (post-contrast), 156 axial T2, and 156 axial FLAIR slices. Performance was evaluated quantitatively using peak signal-to-noise ratio (PSNR) and structural similarity index measure (SSIM).

We compared CM-RED against two DM-based MRI reconstruction methods, DPS~\cite{chung2023dps} and DDS~\cite{chung2024decomposed}, and one CM-based inverse problem solver, CM4IR~\cite{garber2025CM4IR}, using their publicly available implementations. For DPS and DDS, we used pretrained publicly available models trained separately for the knee and brain datasets~\cite{chung2024decomposed}. DPS and DDS were implemented using 1000 and 100 NFEs, respectively. For CM4IR, we used the same pretrained CM models employed in CM-RED and performed reconstruction with 4 NFEs. For all comparison methods, hyperparameters were empirically optimized separately for each acceleration factor, undersampling pattern, and dataset to obtain their best reconstruction performance. 

\begin{table}[!t]
\caption{Quantitative comparison of reconstruction methods on fastMRI~\cite{knoll2020fastmri_dataset-journal} \colorbox{blue!15}{\textbf{knee}} and \colorbox{orange!15}{\textbf{brain}} datasets with $\mathbf{R} = 4$ and $8$ using \textit{1D equidistant} undersampling. Best: \textbf{BOLD}, second-best: \underline{Underlined}.}
% \vspace{-3pt}

\label{tab:results_equidistant}
\begin{center}
\begin{footnotesize}
\setlength{\tabcolsep}{1.3pt}
\renewcommand{\arraystretch}{1.1}
\begin{tabular}{@{}p{1.3cm}cc 
cc!{\color{gray!50}\vrule width 0.4pt}cc|
cc!{\color{gray!50}\vrule width 0.4pt}cc!{\color{gray!50}\vrule width 0.4pt}
cc!{\color{gray!50}\vrule width 0.4pt}cc@{}}

\toprule

\multirow{4}{*}{\textbf{Method}} 
& \multirow{4}{*}{\textbf{NFEs}$\;\downarrow$}
& \multirow{4}{*}{$\mathbf{R}$} &
\multicolumn{2}{>{\columncolor{blue!15}}c}{\textbf{Cor PD}} &
\multicolumn{2}{>{\columncolor{blue!15}}c|}{\textbf{Cor PD-FS}} &
\multicolumn{2}{>{\columncolor{orange!15}}c}{\textbf{Axial T1-Pre}} &
\multicolumn{2}{>{\columncolor{orange!15}}c}{\textbf{Axial T1-Post}} &
\multicolumn{2}{>{\columncolor{orange!15}}c}{\textbf{Axial T2}} &
\multicolumn{2}{>{\columncolor{orange!15}}c}{\textbf{Axial FLAIR}} \\

\cmidrule(lr){4-15}
& & & 
\scriptsize PSNR$\uparrow$ & \scriptsize SSIM$\uparrow$ &
\scriptsize PSNR$\uparrow$ & \scriptsize SSIM$\uparrow$ &
\scriptsize PSNR$\uparrow$ & \scriptsize SSIM$\uparrow$ &
\scriptsize PSNR$\uparrow$ & \scriptsize SSIM$\uparrow$ &
\scriptsize PSNR$\uparrow$ & \scriptsize SSIM$\uparrow$ &
\scriptsize PSNR$\uparrow$ & \scriptsize SSIM$\uparrow$ \\

\toprule

% \multirow{2}{*}{DPS~\cite{chung2023dps}} 
\multirow{2}{*}{\bf DPS} 
& \multirow{2}{*}{1000}
& \scriptsize $\times 4$ 
& 34.82 & 0.886 
& \textbf{32.37} & 0.781 
& 37.70 & 0.928
& 38.25 & 0.932
& \underline{34.49} & 0.911 
& 34.37 & 0.907 \\
& 
& \scriptsize $\times 8$ 
& 31.79 & 0.832
& \underline{30.30} & \underline{0.727}
& 33.75 & 0.896
& 33.89 & 0.899 
& 30.14 & 0.868
& 30.23 & 0.862 \\
\arrayrulecolor{gray!40}\midrule

% \multirow{2}{*}{DDS~\cite{chung2024decomposed}} 
\multirow{2}{*}{\bf DDS}
& \multirow{2}{*}{100}
& \scriptsize $\times 4$ 
& \underline{34.86} & \underline{0.901}
& 31.40 & \underline{0.781}
& \underline{38.24} & \underline{0.958}
& \underline{38.17} & \underline{0.955}
& 34.22 & \underline{0.928}
& \underline{35.29} & \underline{0.936} \\
&
& \scriptsize $\times 8$ 
& \underline{31.63} & \underline{0.837}
& 29.67 & 0.720 
& \underline{34.82} & \underline{0.927}
& 34.63 & \underline{0.923} 
& \underline{30.88} & \underline{0.884}
& \underline{31.35} & \underline{0.887} \\
\arrayrulecolor{gray!40}\midrule

% \multirow{2}{*}{CM4IR~\cite{garber2025CM4IR}} 
\multirow{2}{*}{\bf CM4IR} 
& \multirow{2}{*}{4}
& \scriptsize $\times 4$ 
& 29.96 & 0.766
& 24.32 & 0.528
& 36.83 & 0.918 
& 37.35 & 0.917 
& 31.43 & 0.856 
& 34.47 & 0.915 \\
&
& \scriptsize $\times 8$ 
& 30.38 & 0.786
& 24.34 & 0.519
& 34.70 & 0.912 
& \underline{35.68} & 0.921 
& 29.99 & 0.858 
& 29.49 & 0.842 \\
\arrayrulecolor{gray!40}\midrule

% \multirow{2}{*}{CM-RED {\bf (ours)}} 
\multirow{2}{*}{\bf CM-RED} 
& \multirow{2}{*}{4}
& \scriptsize $\times 4$
& \textbf{36.58} & \textbf{0.927}
& \underline{32.02} & \textbf{0.804}
& \textbf{40.61} & \textbf{0.969} 
& \textbf{40.97} & \textbf{0.969} 
& \textbf{36.13} & \textbf{0.945} 
& \textbf{36.96} & \textbf{0.948} \\
&
& \scriptsize $\times 8$ 
& \textbf{32.94} & \textbf{0.861}
& \textbf{30.46} & \textbf{0.749}
& \textbf{37.02} & \textbf{0.939}
& \textbf{37.06} & \textbf{0.936} 
& \textbf{32.34} & \textbf{0.902} 
& \textbf{33.35} & \textbf{0.908} \\

\arrayrulecolor{black}\bottomrule
\end{tabular}
\end{footnotesize}
\end{center}
\end{table}
\begin{table*}[!t]
\caption{Quantitative comparison of reconstruction methods on fastMRI~\cite{knoll2020fastmri_dataset-journal} \colorbox{blue!15}{\textbf{knee}} and \colorbox{orange!15}{\textbf{brain}} datasets with $\mathbf{R} = 4$ and $8$ using \textit{1D Gaussian} undersampling. Best: \textbf{BOLD}, second-best: \underline{Underlined}.}
\vspace{-3pt}

\label{tab:results_gaussian}
\begin{center}
\begin{footnotesize}
\setlength{\tabcolsep}{1.3pt}
\renewcommand{\arraystretch}{1.1}
\begin{tabular}{@{}p{1.3cm}cc 
cc!{\color{gray!50}\vrule width 0.4pt}cc|
cc!{\color{gray!50}\vrule width 0.4pt}cc!{\color{gray!50}\vrule width 0.4pt}
cc!{\color{gray!50}\vrule width 0.4pt}cc@{}}

\toprule

\multirow{4}{*}{\textbf{Method}} 
& \multirow{4}{*}{\textbf{NFEs}$\;\downarrow$}
& \multirow{4}{*}{$\mathbf{R}$} &
\multicolumn{2}{>{\columncolor{blue!15}}c}{\textbf{Cor PD}} &
\multicolumn{2}{>{\columncolor{blue!15}}c|}{\textbf{Cor PD-FS}} &
\multicolumn{2}{>{\columncolor{orange!15}}c}{\textbf{Axial T1-Pre}} &
\multicolumn{2}{>{\columncolor{orange!15}}c}{\textbf{Axial T1-Post}} &
\multicolumn{2}{>{\columncolor{orange!15}}c}{\textbf{Axial T2}} &
\multicolumn{2}{>{\columncolor{orange!15}}c}{\textbf{Axial FLAIR}} \\

\cmidrule(lr){4-15}
& & & 
\scriptsize PSNR$\uparrow$ & \scriptsize SSIM$\uparrow$ &
\scriptsize PSNR$\uparrow$ & \scriptsize SSIM$\uparrow$ &
\scriptsize PSNR$\uparrow$ & \scriptsize SSIM$\uparrow$ &
\scriptsize PSNR$\uparrow$ & \scriptsize SSIM$\uparrow$ &
\scriptsize PSNR$\uparrow$ & \scriptsize SSIM$\uparrow$ &
\scriptsize PSNR$\uparrow$ & \scriptsize SSIM$\uparrow$ \\

\toprule

% \multirow{2}{*}{DPS~\cite{chung2023dps}} 
\multirow{2}{*}{\bf DPS}
& \multirow{2}{*}{1000}
& \scriptsize $\times 4$ 
& \underline{35.19} & 0.894
& \underline{32.81} & 0.795
& 35.36 & 0.915
& 35.88 & 0.918
& 33.37 & 0.903
& 32.92 & 0.901 \\
% \addlinespace[1pt]
& 
& \scriptsize $\times 8$ 
& \underline{33.64} & 0.862
& \textbf{31.65} & \underline{0.757}
& 33.87 & 0.906
& 34.16 & 0.909
& \underline{32.22} & 0.893
& 31.92 & 0.887 \\
\arrayrulecolor{gray!40}\midrule

% \multirow{2}{*}{DDS~\cite{chung2024decomposed}} 
\multirow{2}{*}{\bf DDS}
& \multirow{2}{*}{100}
& \scriptsize $\times 4$ 
% & 34.93 & 0.909 % zeta = 0.5 PD
% & 31.64 & 0.784 % zeta = 2.5 PDFS
& 35.10 & \underline{0.909} % zeta = 0.3 PD
& 31.94 & \underline{0.796} % zeta = 2.0 PDFS
& \underline{37.77} & \underline{0.956}
& \underline{38.00} & \underline{0.956} 
& \underline{34.15} & \underline{0.928}
& \underline{35.47} & \underline{0.940} \\
% \addlinespace[1pt]
&
& \scriptsize $\times 8$ 
% & 32.90 & 0.869  % zeta = 0.5 PD
% & 30.16 & 0.734 % zeta = 2.5 PDFS
& 33.14 & \underline{0.871} % zeta = 0.3 PD
& 30.47 & 0.745 % zeta = 2.0 PDFS
& \underline{35.63} & \underline{0.939}
& 35.68 & \underline{0.936}
& 32.19 & \underline{0.903}
& \underline{33.14} & \underline{0.912} \\
% \addlinespace[1pt]
\arrayrulecolor{gray!40}\midrule

% \multirow{2}{*}{CM4IR~\cite{garber2025CM4IR}} 
\multirow{2}{*}{\bf CM4IR} 
& \multirow{2}{*}{4}
& \scriptsize $\times 4$ 
& 32.71 & 0.839
& 26.58 & 0.598
& 36.62 & 0.927 
& 37.63 & 0.934 
& 32.31 & 0.879 
& 33.33 & 0.900 \\
% \addlinespace[1pt]
&
& \scriptsize $\times 8$ 
& 32.38 & 0.833
& 25.76 & 0.561
& \underline{36.04} & 0.927 
& \underline{36.40} & 0.928 
& 31.55 & 0.885 
& 31.92 & 0.878 \\
% \addlinespace[1pt]
\arrayrulecolor{gray!40}\midrule

% \multirow{2}{*}{CM-RED {\bf (ours)}} 
\multirow{2}{*}{\bf CM-RED} 
& \multirow{2}{*}{4}
& \scriptsize $\times 4$
& \textbf{36.58} & \textbf{0.930}
& \textbf{33.24} & \textbf{0.838}
& \textbf{40.06} & \textbf{0.969} 
& \textbf{40.46} & \textbf{0.968} 
& \textbf{35.51} & \textbf{0.943} 
& \textbf{37.41} & \textbf{0.953} \\
% \addlinespace[1pt]
&
& \scriptsize $\times 8$ 
& \textbf{34.15} & \textbf{0.886}
& \underline{31.19} & \textbf{0.770}
& \textbf{37.40} & \textbf{0.945} 
& \textbf{37.37} & \textbf{0.938} 
& \textbf{33.06} & \textbf{0.911} 
& \textbf{34.71} & \textbf{0.922} \\

\arrayrulecolor{black}\bottomrule
\end{tabular}

\end{footnotesize}
\end{center}
\end{table*}

\section{Experiments and Results}
\label{sec:results}

\subsection{Reconstruction Results}

Figs.~\ref{fig:results_equi} and~\ref{fig:results_gauss} show representative reconstructions for equidistant and Gaussian undersampling, respectively. DPS exhibits reduced sharpness and residual aliasing in several cases, while DDS improves sharpness but does not fully suppress the artifacts. CM4IR shows noticeable noise amplification due to its use of a pseudo-inverse back-projection step, degrading the overall visual quality. In contrast, CM-RED more consistently suppresses residual artifacts while preserving anatomical detail across the evaluated settings. This is particularly visible in the axial T1-pre example in Fig.~\ref{fig:results_equi}, where the lateral ventricles are more clearly delineated.

Tab.~\ref{tab:results_equidistant} and Tab.~\ref{tab:results_gaussian} summarize the quantitative reconstruction results for equidistant and Gaussian undersampling, respectively, at $R=4$ and $R=8$. Consistent with the qualitative comparisons, CM-RED achieves the highest PSNR and SSIM in nearly all configurations despite using only 4 NFEs. The only exceptions are the PSNR values for coronal PD-FS at $R=4$ with equidistant undersampling and at $R=8$ with Gaussian undersampling, where CM-RED achieves the second-best performance. Notably, DPS and DDS require substantially more NFEs (1000 and 100, respectively) yet still underperform compared to CM-RED. Although CM4IR also operates with 4 NFEs, its reliance on a pseudo-inverse back-projection step limits its applicability to MRI forward models, resulting in inferior reconstruction performance.

\begin{figure}[t]
    \centering
    \begin{subfigure}[t]{0.48\linewidth}
        \centering
        \includegraphics[width=\linewidth]{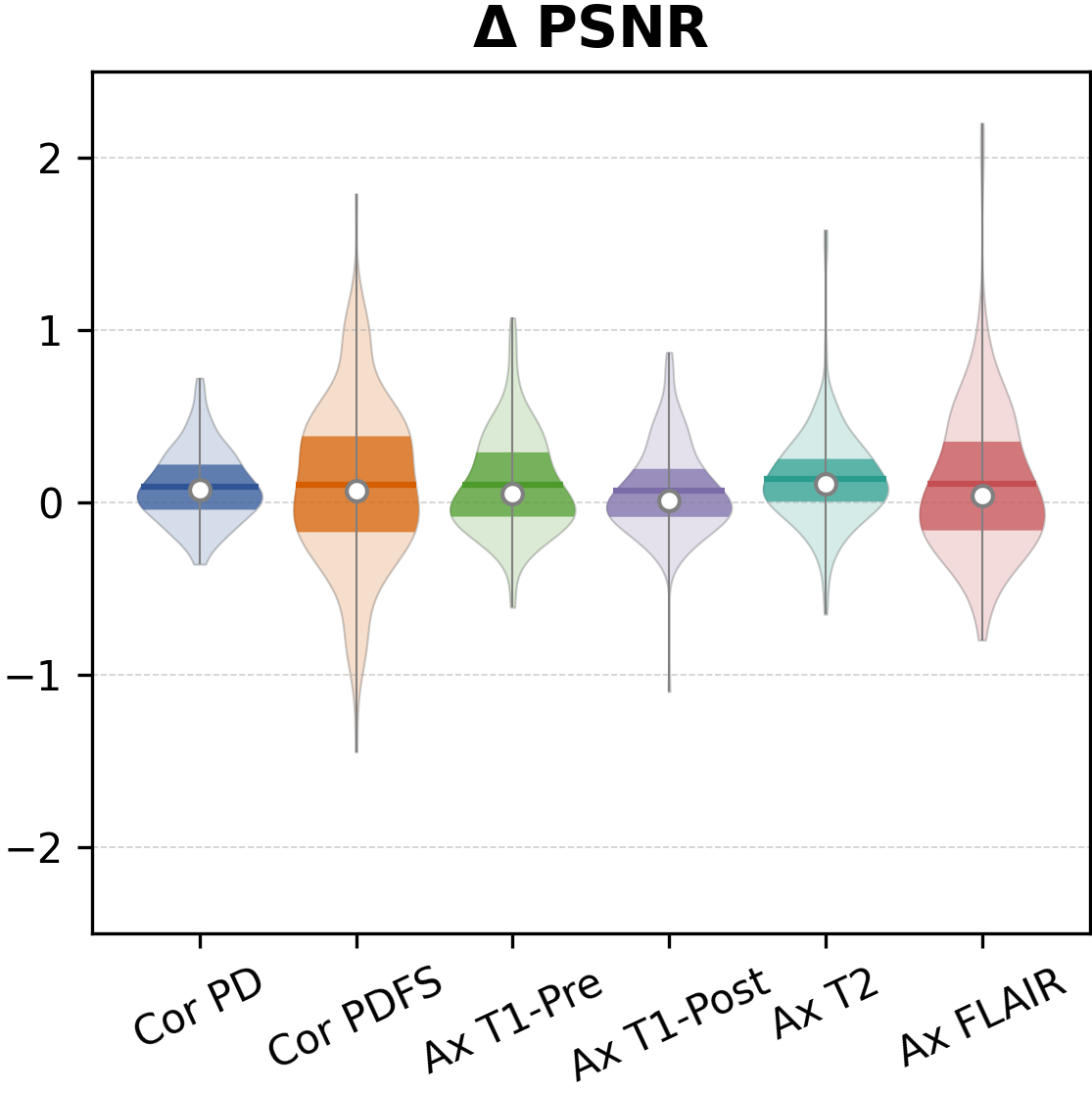}
    \end{subfigure}
    \hfill
    \begin{subfigure}[t]{0.48\linewidth}
        \centering
        \includegraphics[width=\linewidth]{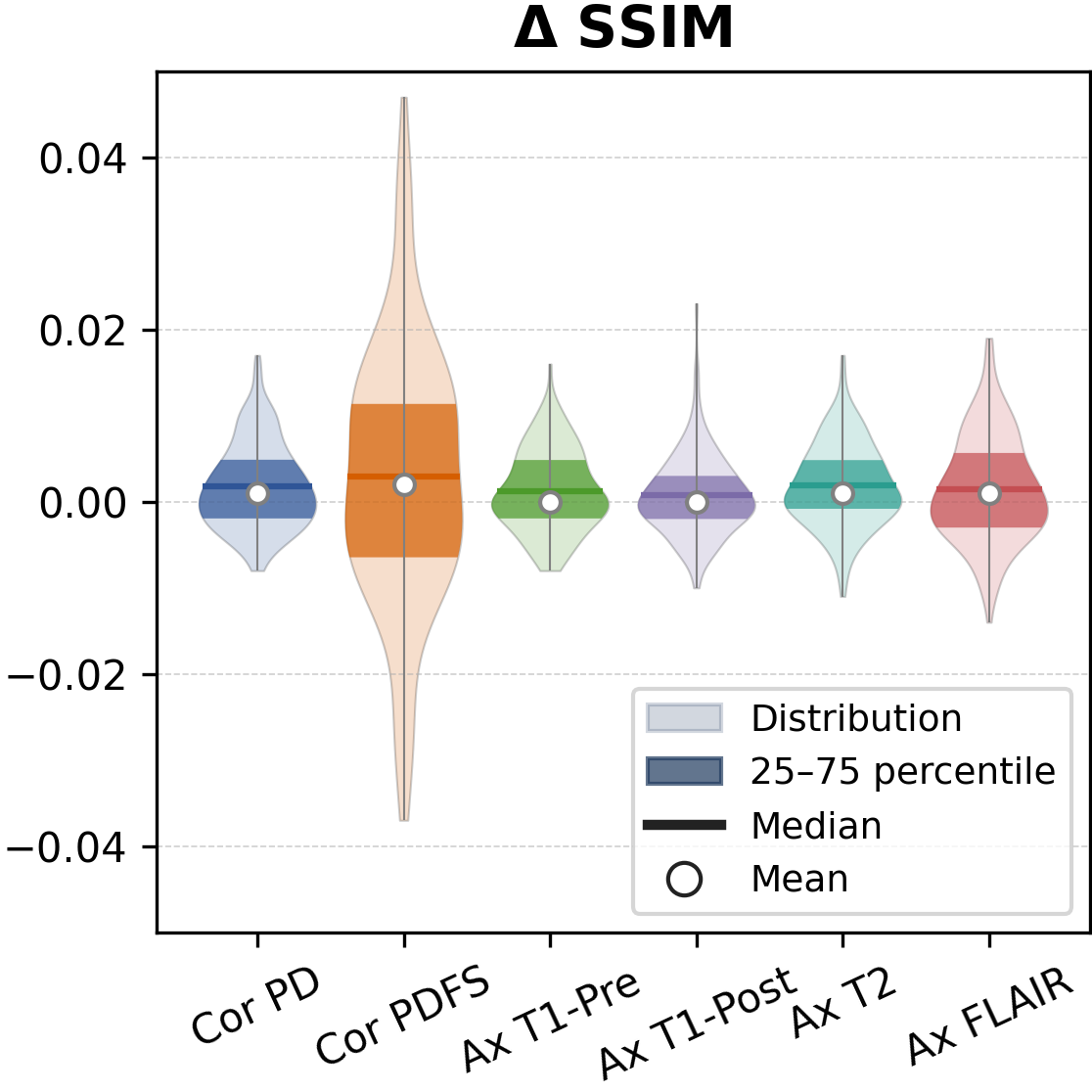}
    \end{subfigure}
    \caption{Hyperparameter sensitivity analysis for CM-RED with $R=8$ Gaussian undersampling. Violin plots show the per-slice difference in PSNR (left) and SSIM (right) between the nominal configuration and reconstructions obtained with randomly perturbed hyperparameters.}
    \label{fig:hyperparam_sense}
\end{figure}

\subsection{Hyperparameter Sensitivity}
To evaluate the robustness of CM-RED to hyperparameter selection, we consider a representative setting with $R=8$ Gaussian undersampling. Starting from the nominal hyperparameter configuration reported in Tab.~\ref{tab:hyperparams}, we introduce independent random perturbations to each hyperparameter during inference. For each slice, every hyperparameter is scaled by an independent factor sampled from $\mathcal{U}(0.9,\,1.1)$, corresponding to $\pm10\%$ variations. We then compute the per-slice PSNR and SSIM differences between the nominal and perturbed reconstructions. As shown in Fig.~\ref{fig:hyperparam_sense}, the distributions remain centered near zero across all six datasets, with 25--75 percentile ranges within approximately $\pm0.2$ dB in PSNR and $\pm0.01$ in SSIM. These results indicate robustness to moderate perturbations around the nominal configuration.

\subsection{Ablation Studies}
\label{sec:ablation_studies}

Our first ablation study assessed the individual and combined effects of two CM-RED components: controlled noise injection in the CM update and momentum acceleration after the data fidelity update. Four variants were evaluated: (i) without noise injection or momentum, (ii) with momentum only, (iii) with noise injection only, and (iv) with both. Experiments were conducted using 4 NFEs on coronal PD knee and axial T2 brain data with $R=4$ equidistant undersampling. The results in Table~\ref{tab:ablation_momentum_noise_inj} indicate that both components improve reconstruction quality, and the variant combining them consistently yields the highest performance. This demonstrates that momentum and noise injection provide complementary benefits across different anatomies and contrasts.

\begin{table}[b]%[15]{r}{0.65\textwidth}
% \captionsetup{type=table}
\caption{PSNR/SSIM on coronal PD knee and axial T2 brain R=4 equidistant undersampling with/without momentum and noise injection using 4 NFEs.
}
\label{tab:ablation_momentum_noise_inj}
\begin{center}
\begin{small}
\renewcommand{\arraystretch}{1.0}
\setlength{\tabcolsep}{10pt}
\begin{tabular}{cccc}
\toprule
\multicolumn{1}{c}{\textbf{Noise Inj.}} &
\multicolumn{1}{c}{\textbf{Momentum}} &
\multicolumn{1}{c}{\textbf{Cor PD Knee}} &
\multicolumn{1}{c}{\textbf{Ax T2 Brain}}\\
\midrule

\xmark & \xmark
    & 35.59 / 0.917 
    & 34.96 / 0.936 \\
\arrayrulecolor{gray!40}\midrule

\xmark & \cmark
    & \underline{36.09} / \underline{0.923} 
    & \underline{35.25} / \underline{0.937} \\
\arrayrulecolor{gray!40}\midrule

\cmark & \xmark 
    & 35.71 / 0.917
    & 35.23 / \underline{0.937} \\
\arrayrulecolor{gray!40}\midrule

\cmark & \cmark 
    & \textbf{36.58} / \textbf{0.927}
    & \textbf{36.13} / \textbf{0.945} \\
\arrayrulecolor{black}\bottomrule
\end{tabular}
\end{small}
\end{center}
\vspace{-1.5ex}
\end{table}

\begin{figure}[t]
    \centering
    \begin{subfigure}[t]{0.49\linewidth}
        \centering
        \includegraphics[width=\linewidth]{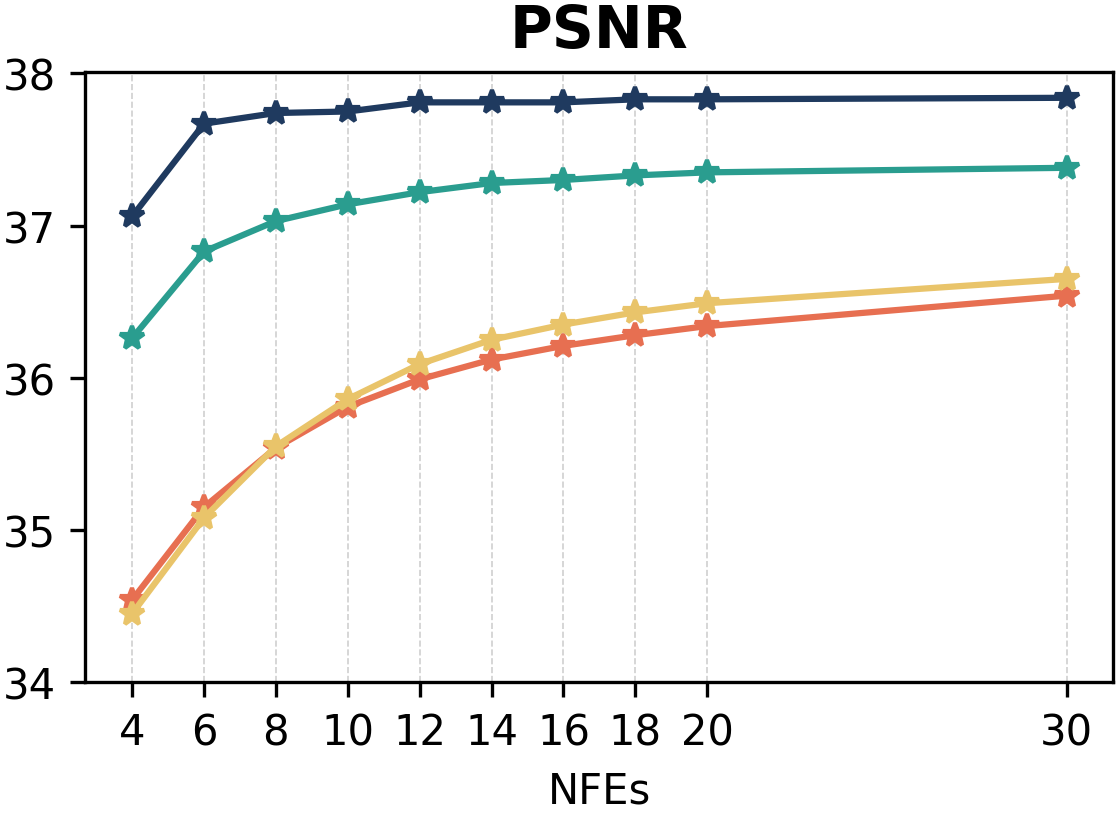}
    \end{subfigure}
    \hfill
    \begin{subfigure}[t]{0.49\linewidth}
        \centering
        \includegraphics[width=\linewidth]{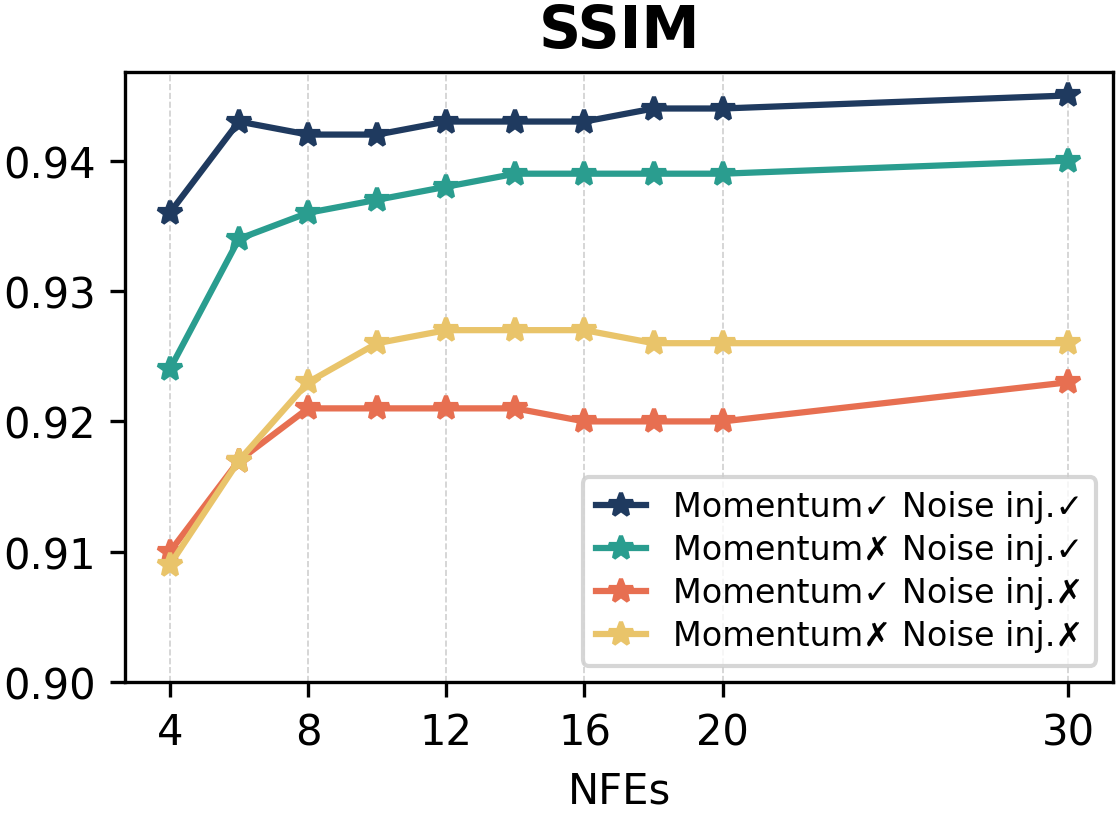}
    \end{subfigure}
    \caption{Effect of noise injection and momentum on CM-RED convergence for fastMRI axial T1-post brain data with R=8 equidistant undersampling. PSNR and SSIM are reported between 4 to 30 NFEs.}
    \label{fig:ablation}
\end{figure}

Our second ablation study further evaluated how these components affect reconstruction quality with increasing NFEs. Fig.~\ref{fig:ablation} shows the performance of the same four variants on axial T1-post brain data with $R=8$ equidistant undersampling for 4--30 NFEs. Across the evaluated NFE range, the configuration using both noise injection and momentum consistently achieved the highest PSNR and SSIM. In fact, even at NFE = 6, this configuration already reaches performance comparable to the best results of the other variants at their convergence point. As expected, momentum and noise injection help avoid shallow local minima, allowing the reconstruction to converge to a substantially better solution. Importantly, noise injection and momentum do not introduce a separate reconstruction objective; instead, they modify the update dynamics within the same CM-RED reconstruction formulation. 

\section{Discussion}
\label{sec:discussion}

This work demonstrates that CMs can serve as computationally efficient generative priors for accelerated MRI reconstruction. By embedding a pretrained CM into a RED-APG-inspired reconstruction scheme, CM-RED retains the fast few-step sampling behavior of CMs while enforcing consistency with acquired multi-coil k-space measurements. Across knee and brain MRI datasets, contrast weightings, acceleration factors, and undersampling patterns.

In our framework, the CM serves as the denoising prior during iterative reconstruction. While DM-based priors estimate clean images through approximate likelihood updates from noisy states, CM-RED directly uses the CM clean-image estimate within an explicit data-fidelity step. This avoids repeatedly traversing a long diffusion trajectory and enables few-step reconstruction. A related line of work in the broader inverse problems community has shown increasing interest in flow matching models~\cite{liu2023flow}, which leverage rectified flow-based generative priors as strong priors for inverse problems~\cite{patel2025flowchef, kim2025flowdps} and enable high-quality few-step generation~\cite{shi2026nullflow}. However, these approaches have primarily focused on latent-space generative models. To the best of our knowledge, no publicly available flow-matching generative model has been trained specifically on MRI data, which has unfortunately precluded a direct comparison with CM-RED.

A central design choice in CM-RED is the joint use of noise injection and momentum acceleration within the iterative updates. Controlled noise injection enables the CM to operate at non-negligible denoising levels, which empirically leads to more effective prior-driven updates under a limited NFE budget. Momentum acceleration, inherited from RED-APG, further improves the efficiency of these updates. Together, these components improve the effectiveness of the iterative updates under a limited NFE budget. The ablation studies support this interpretation: noise injection and momentum each improve reconstruction quality individually, while their combination provides the strongest performance across the evaluated settings.

Although CM-RED includes several hand-tuned reconstruction hyperparameters, the proposed implementation was shown to be robust to moderate variations around the nominal configuration. This is important for practical MRI reconstruction, where exhaustive protocol-specific tuning is undesirable. In our experiments, the same hyperparameter configuration was used across anatomies, contrast weightings, and undersampling patterns, with separate settings only for different acceleration factors. The sensitivity analysis further showed that moderate perturbations of the selected parameters led to limited changes in PSNR and SSIM, suggesting that CM-RED operates reliably within a reasonably broad parameter range.

Compared with existing CM-based inverse problem solvers such as CM4IR, the main distinction of CM-RED lies in the form of the data fidelity update. Back-projection-based approaches can be effective for some image restoration problems, but in accelerated multi-coil MRI where measurements are inherently noisy, back-projection steps may amplify measurement noise. CM-RED addresses these limitations by combining the strong generative prior of CMs with a quadratic penalized data fidelity update based on the multi-coil MRI forward model. 
This design allows the reconstruction to benefit from the learned CM prior while explicitly enforcing consistency with the acquired k-space measurements. Empirically, this balance improves reconstruction quality across both fastMRI knee and brain datasets.

It is also important to distinguish CM-RED from PD-DL reconstruction methods. PD-DL methods can achieve excellent performance, particularly when trained and evaluated under matched acquisition protocols~\cite{hammernik2018VarNet, aggarwal2019MoDL, knoll2020deep-survey, hammernik2023SPM, yaman2022zeroshot}, but they are typically optimized for specific sampling patterns, acceleration factors, anatomies, and contrast weightings. CM-RED is positioned differently: it uses a pretrained generative prior at test time and combines it with the measurement model without end-to-end retraining for each undersampling pattern. Thus, the goal of this work is not to replace PD-DL reconstruction networks, but to demonstrate that CMs can be used as efficient and flexible generative priors for MRI inverse problems.

The results also suggest several directions for future work. Although CM-RED requires only a few NFEs at inference, training the underlying DM and distilling it into a CM remain computationally demanding. Training CM priors on broader MRI datasets and pathology-rich cohorts may further improve their generalizability and practical utility.  Similarly, evaluating CM-RED on non-Cartesian scans and patient cohorts may help assess its robustness in other acquisition settings. Another important direction is reducing the dependence on fully-sampled training data, which were used for training the generative priors in this work. However, fully-sampled datasets are not always available in clinical MRI. Recent unsupervised and measurement-domain generative learning approaches for DMs, including AmbientGAN-style training and ambient diffusion/posterior sampling methods, provide a potential path toward learning MRI priors directly from undersampled or corrupted measurements~\cite{aali2025ambient-diff} and a method for training/distilling CMs in a similar setting may be valuable for practical purposes.

\section{Conclusion}
\label{sec:concusion}
We introduced CM-RED, a few-step MRI reconstruction framework that uses pretrained consistency models as a learned proximal operator. CM-RED achieved consistently strong reconstruction quality across knee and brain MRI experiments, while requiring only four NFEs, outperforming existing DM- and CM-based solvers in nearly all settings. Overall, CM-RED provides a practical, computationally efficient, and reliable generative-model-based reconstruction strategy for accelerated MRI.

\section*{Acknowledgment}
This work was partially supported by NIH R01EB032830, NIH R01HL179616 and NIH P41EB027061.

\bibliographystyle{johd}
\bibliography{refs}

\end{document}